\documentclass[fleqn,usenatbib]{mnras}

\usepackage{newtxtext,newtxmath}

\usepackage[T1]{fontenc}
\usepackage{hyperref}

\DeclareRobustCommand{\VAN}[3]{#2}
\let\VANthebibliography\thebibliography
\def\thebibliography{\DeclareRobustCommand{\VAN}[3]{##3}\VANthebibliography}

\usepackage{graphicx}	
\usepackage{amsmath}	
\usepackage{float}
\usepackage{booktabs}

\newcommand{\Msun}{\,{\rm M_\odot}}
\newcommand{\Mblack}{M_\bullet}
\newcommand{\dist}{\, {\rm ckpc} \, h^{-1}}

\def\hmpc{h^{-1}{\rm Mpc}}
\def\hmsun{{h^{-1} \rm M_{\odot}}}

\def\hkpc{h^{-1}\, {\rm kpc}}
\def\astrid{\texttt{ASTRID} }

\usepackage{xcolor}

\title[Wandering IMBHs in ASTRID]{Orbital and Radiative Properties of Wandering Intermediate-Mass Black Holes in the ASTRID Simulation}

\author[E. Weller et al.]{Emma Jane Weller$^{1}$\thanks{emmaweller@college.harvard.edu},
Fabio Pacucci$^{1,2}$\thanks{fabio.pacucci@cfa.harvard.edu}, Yueying Ni$^{1,3}$, Nianyi Chen$^{3}$, Tiziana Di Matteo$^{3,4}$, \newauthor Magdalena Siwek$^{1}$, Lars Hernquist$^{1}$
\\
$^{1}$Center for Astrophysics $\vert$ Harvard \& Smithsonian, Cambridge, MA 02138, USA\\
$^{2}$Black Hole Initiative, Harvard University,
Cambridge, MA 02138, USA\\
$^{3}$McWilliams Center for Cosmology, Department of Physics, Carnegie Mellon University, Pittsburgh, PA 15213\\
$^{4}$NSF AI Planning Institute for Physics of the Future, Carnegie Mellon University, Pittsburgh, PA 15213, USA
}

\date{\today}

\pubyear{2022}

\begin{document}
\label{firstpage}
\pagerange{\pageref{firstpage}--\pageref{lastpage}}
\maketitle

\begin{abstract}
Intermediate-Mass Black Holes (IMBHs) of $10^3-10^6\Msun$ are commonly found at the center of dwarf galaxies. Simulations and observations convincingly show that a sizable population of IMBHs could wander off-center in galaxies. We use the cosmological simulation \astrid to study the orbital and radiative properties of wandering IMBHs in massive galaxies at $z\sim3$.
We find that this population of black holes has large orbital inclinations ($60^\circ\pm22^\circ$) with respect to the principal plane of the host. The eccentricity of their orbits is also significant ($0.6\pm0.2$) and decreases with time. Wandering IMBHs undergo spikes of accretion activity around the pericenter of their orbits, with rates $10^{-3}-10^{-5}$ times the Eddington rate and a median accretion duty cycle of $\sim 12\%$. Their typical spectral energy distribution peaks in the infrared at $\sim 11 \, \mu \rm m$ rest-frame. Assuming a standard value of $10\%$ for the matter-to-energy radiative efficiency, IMBHs reach $2-10$ keV X-ray luminosities $>10^{37} \, \mathrm{erg\,s^{-1}}$ for $\sim10\%$ of the time. This luminosity corresponds to fluxes $>10^{-15} \, \mathrm{erg \, s^{-1} \, cm^{-2}}$ within $10$ Mpc. They could be challenging to detect because of competing emissions from X-ray binaries and the interstellar medium. X-ray luminosities $> 10^{41} \, \mathrm{erg \, s^{-1}}$, in the hyper-luminous X-ray sources (HLXs) regime, are reached by $\sim 7\%$ of the IMBHs. These findings suggest that HLXs are a small subset of the wandering IMBH population, which is characterized by luminosities $10^3-10^4$ times fainter. Dedicated surveys are needed to assess the demographics of this missing population of black holes.
\end{abstract}




\section{Introduction} \label{sec:intro}

Black holes (BHs) cover a wide mass range and are commonly detected throughout the known Universe. The community has extensively investigated two populations at the extremes of the BH mass range. Stellar-mass black holes typically have masses $\lesssim 10^3 \Msun$ and are found throughout galaxies. For example, the Milky Way (MW) may host $\sim 10^8$ stellar-mass BHs (e.g., \citealt{Elbert_2018}). On the other hand, supermassive black holes (SMBHs) have masses $\gtrsim 10^6 \Msun$ and are usually found at the center of massive galaxies. The MW has a central SMBH with a mass of $\sim 4 \times 10^6 \Msun$ (e.g., \citealt{Ghez_2008, Genzel_2010}), which was recently imaged by the Event Horizon Telescope Collaboration \citep{EHT_2022}.

Intermediate-mass black holes (IMBHs), which we define in the mass range $10^3 \Msun \lesssim \Mblack \lesssim 10^6 \Msun$, bridge the gap between stellar-mass and supermassive black holes. A growing population of IMBHs has been identified in dwarf galaxies using several methods, including X-rays, broad and narrow emission lines, infrared emission, nuclear variability, radio emission, and masers (see, e.g., the review by \citealt{Greene_2020_IMBH} and references therein). Recent simulations \citep{Bellovary_2019} and observations \citep{Reines_2020} show that approximately half of massive black holes ($\sim 10^3$ to $\sim 10^7 \Msun$) in dwarf galaxies are off-center, mainly due to galaxy-galaxy mergers \citep{Bellovary_2021}. Recently, \cite{Pacucci_2021} predicted that $5 \%$ to $22 \%$ of dwarf galaxies are active, i.e., they host a central IMBH accreting at least at $10\%$ of the Eddington rate.

\begin{figure*}
	\includegraphics[width=2\columnwidth]{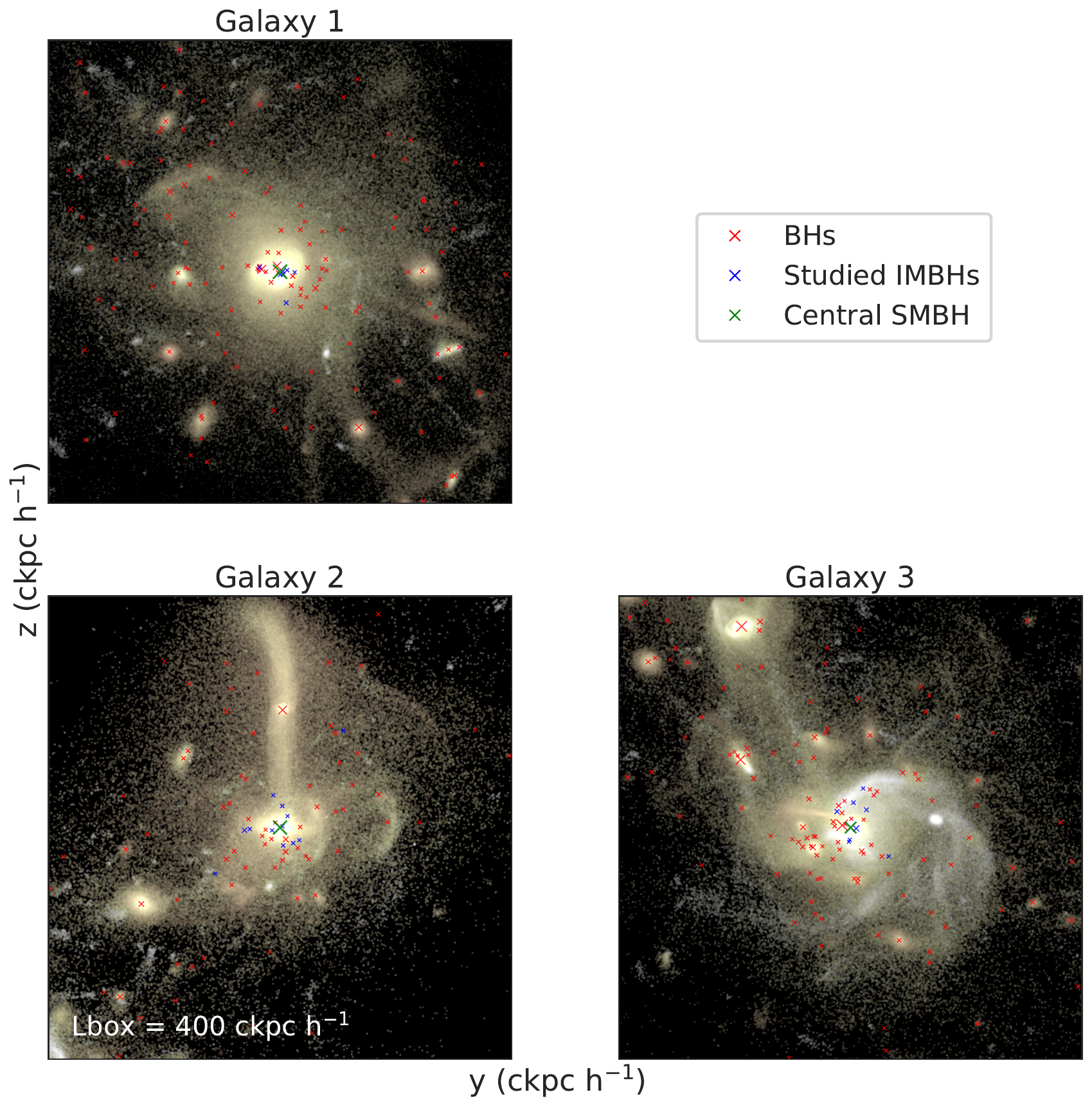}
    \caption{Visualization of the three investigated galaxies at $z=3$. The background shows the stellar density field (in the y-z plane) centered at the SMBH, in a $400 \dist$ region. The marked elements are labeled legend and the marker sizes scale with the BH masses.}
    \label{fig:galaxies}
\end{figure*}

\begin{table*}
\centering
\caption{Properties at $z=3$ of the three galaxies shown in Fig. \ref{fig:galaxies}. Figures are rounded to one decimal.}
\label{tab:galaxy_properties}
\begin{tabular}{cccccccc}
\hline
\textbf{Galaxy} & \textbf{Total subhalo} & \textbf{Half-mass} & \textbf{Stellar} & \textbf{Stellar half-mass} & \textbf{Number of} & \textbf{Number of} & \textbf{Mass of} \\
& \textbf{mass} ($\Msun$) & \textbf{radius} ($\dist$) & \textbf{mass} ($\Msun$) & \textbf{radius} ($\dist$) & \textbf{BHs in subhalo} & \textbf{studied IMBHs} & \textbf{central SMBH} ($\Msun$) \\ \hline
$1$ & $3.4 \times 10^{13}$ & $94.1$ & $8.1 \times 10^{11}$ & $0.7$ & $103$ & $8$ & $2.2 \times 10^9$ \\
$2$ & $2.7 \times 10^{13}$ & $128.7$ & $4.3 \times 10^{11}$ & $0.6$ & $62$ & $12$ & $1.9 \times 10^9$ \\
$3$ & $2.4 \times 10^{13}$ & $75.2$ & $4.0 \times 10^{11}$ & $1.7$ & $89$ & $8$ & $4.3 \times 10^8$ \\
\hline
\end{tabular}
\end{table*}

Through galaxy-galaxy mergers, IMBHs could wander in galaxies like ours. The MW is estimated to have experienced $15 \pm 3$ mergers with galaxies with stellar masses $\gtrsim 4.5 \times 10^6 \Msun$ \citep{Kruijssen_2020}. Recent studies have begun to investigate this undetected population of wandering BHs (see, e.g., \citealt{Bellovary_2010, Gonzalez_2018, Tremmel_2018, Greene_2021, Ricarte_2021a, Ricarte_2021b, Seepaul_2022, Weller_2022, TDM_2022}).

Other proposed methods for IMBH formation include runaway mergers in globular star clusters, hyper-Eddington accretion onto stellar-mass BHs, supra-exponential accretion on seed BHs, and direct collapse of hypermassive quasi-stars (see, e.g., 
\citealt{Governato_1994, Volonteri_2003, HB_2008, Fragione_2018a, Greene_2020_IMBH, Pacucci_2020} and references therein).

This study investigates the orbital and radiative properties of wandering IMBHs in massive $z\sim 3$ galaxies using the cosmological simulation \astrid (\citealt{Bird_2022, Ni_2022, Chen_2022}). We confirm some of our previous findings reported by \cite{Weller_2022}, which used the Illustris TNG50 simulation \citep{Nelson_2019_Illustris, Nelson_2019_TNG50, Pillepich_2019}, but primarily focus on new results. In the following \S \ref{sec:simulation}, we describe the \astrid simulation in detail and explain our process for selecting BHs for our sample. In \S \ref{sec:orbits}, we study some orbital properties of IMBHs, including orbital inclination and eccentricity, along with their evolution. In \S \ref{sec:accretion}, we describe their accretion and radiative properties and investigate their detectability. Finally, in \S \ref{sec:conclusions} we summarize our work and discuss its implications.

\section{Methodology} \label{sec:simulation}

\subsection{ASTRID — The Cosmological Simulation} \label{sec:ASTRID}

\astrid is a cosmological hydrodynamical simulation performed using a new version of the \texttt{MP-Gadget} simulation code. 

\astrid contains $5500^3$ cold dark matter (DM) particles in a $250 \hmpc$ side box and an initially equal number of SPH hydrodynamic mass elements.
The cosmological parameters used in \astrid are from \cite{Planck}, with $\Omega_0=0.3089$, $\Omega_\Lambda=0.6911$, $\Omega_{\rm b}=0.0486$, $\sigma_8=0.82$, $h=0.6774$, $A_s = 2.142 \times 10^{-9}$, $n_s=0.9667$. 
The gravitational softening length is $\epsilon_{\rm g} = 1.5 \hkpc$ for both DM and gas particles.
\astrid achieves a dark matter particle mass resolution of $9.6\times 10^6 \Msun$ and $M_{\rm gas} = 1.3 \times 10^6 \Msun$ in the initial conditions.
The \astrid simulation has reached $z=1.7$ and plans to reach $z=1$.  

\astrid implements a variety of sub-grid models for physics governing the formation of galaxies and SMBHs and their associated supernova and AGN feedback, inhomogeneous hydrogen, helium reionization, and the effect of massive neutrinos.
Here we briefly summarize the physical models used in the simulation and refer the readers to the introductory paper \cite{Ni_2022,Bird_2022} for more detailed descriptions.
 
In {\texttt{ASTRID}}, gas is allowed to cool via primordial radiative cooling \citep{Katz1996ApJS..105...19K} and via metal line cooling, with the gas and stellar metallicities traced following \cite{Vogelsberger:2014}.
Patchy reionization of hydrogen is performed with a spatially varying ultra-violet background using a semi-analytic method based on radiative transfer simulations \citep{Battaglia2013ApJ...776...81B}. 
For the ionized regions, we applied the ionizing ultra-violet background from \cite{FG2020} and gas self-shielding following \cite{Rahmati:2013}. 
We implement star formation based on the multi-phase stellar formation model in \cite{SH03}, which accounts for the effects of molecular hydrogen~\citep{Krumholz2011ApJ...729...36K}.
Type II supernova wind feedback is included following \cite{Okamoto2010}, assuming wind speeds proportional to the local one-dimensional dark matter velocity dispersion.

Subgrid models for SMBH applied in \astrid can be summarized as follows.
SMBHs are represented by particles that can accrete gas, merge and apply feedback to their baryonic surroundings.
We use halo-based seeding where the BHs are seeded in haloes with $M_{\rm halo,FOF} > 5 \times 10^9 \hmsun$ and $M_{\rm *,FOF} > 2 \times 10^6 \hmsun$. 
Seed masses are stochastically drawn from a power-law probability distribution, with a mass between $3\times10^{4} \hmsun$ and $3\times10^{5} \hmsun$ and power-law index $n = -1$.

The gas accretion rate onto the BH is estimated via a Bondi-Hoyle-Lyttleton-like prescription \citep{DSH2005}:
\begin{equation}
\centering
\label{equation:Bondi}
    \dot{M}_{\rm B} = \frac{4 \pi \alpha G^2 M_{\rm BH}^2 \rho}{(c^2_s+v_{\rm rel}^2)^{3/2}}
\end{equation}
where $c_s$ and $\rho$ are the local sound speed and density of gas, $v_{\rm rel}$ is the relative velocity of the BH with respect to the nearby gas and $\alpha = 100$ is a dimensionless fudge parameter to account for the underestimation of the accretion rate due to the unresolved cold and hot phase of the subgrid interstellar medium in the surrounding.
We allow for short periods of super-Eddington accretion in the simulation but limit the accretion rate to two times the Eddington accretion rate.
The BH radiates with a bolometric luminosity $L_{\rm bol}$ proportional to the accretion rate $\dot{M}_\bullet$, with a mass-to-energy conversion efficiency $\eta=0.1$ in an accretion disk according to \cite{Shakura1973}.
\begin{equation}
\centering
\label{equation:Lbol}
    L_{\rm Bol} = \eta \dot{M}_{\rm BH} c^2
\end{equation}
A percentage of 5\% of the radiated energy is coupled to the surrounding gas as the AGN feedback.
The feedback from SMBHs includes what is often referred to as quasar-mode or thermal feedback, as well as kinetic feedback.

The dynamics of the SMBHs are modeled with a newly developed (sub-grid) dynamical friction model \citep{Tremmel2015,Chen2021} to replace the original implementation that directly repositioned the BHs to the minimum local potential.
This model provides an improved treatment for calculating BH trajectories and velocities.
However, BH particles near the host's center are subject to spurious dynamical heating from other particles significantly larger than the galactic component they represent. 
Hence, this effect may reflect in the initial stages of the BH's orbit, when it resides in the central region of the host. 
Two BHs merge if their separation is within two times the spatial resolution $2\epsilon_g$, once their kinetic energy is dissipated by dynamical friction and they are gravitationally bound to each other.
In order to reduce the noisy gravitational forces (dynamical heating) acting on the small seed mass black holes, we use another BH mass tracer, the dynamical mass $M_{\rm dyn}$, to account for the force calculation of BH (including the gravitational force and dynamical friction).
When a new BH is seeded, we initialize the corresponding $M_{\rm dyn} = M_{\rm dyn,seed} = 10^7 \hmsun$, which is about $1.5 M_{\rm DM}$. 
\cite{Chen2021} showed that this alleviates dynamic heating and stabilizes the BH motion in the early growth phase.
$M_{\rm dyn}$ is kept at its seeding value until $M_{\rm BH}>M_{\rm dyn,seed}$.
After that, $M_{\rm dyn}$ grows following the BH mass accretion.

The validation of the dynamical friction model in cosmological simulations is described in \cite{Chen2021}.

\astrid provides ``BH details'' files that log the BH evolution at high time resolution and allow detailed analysis of the BH orbital trajectories and light curves. 
Among the BH snapshots used in our analysis, the median time spacing is $\sim 4.8 \times 10^{-5} \rm \, Gyr$.
We used the lowest redshift data available, and we studied most of our BHs up to $\sim 2.8 \rm \, Gyr$ of cosmic time (see \S \ref{sec:infall}).

\begin{figure}
	\includegraphics[width=\columnwidth]{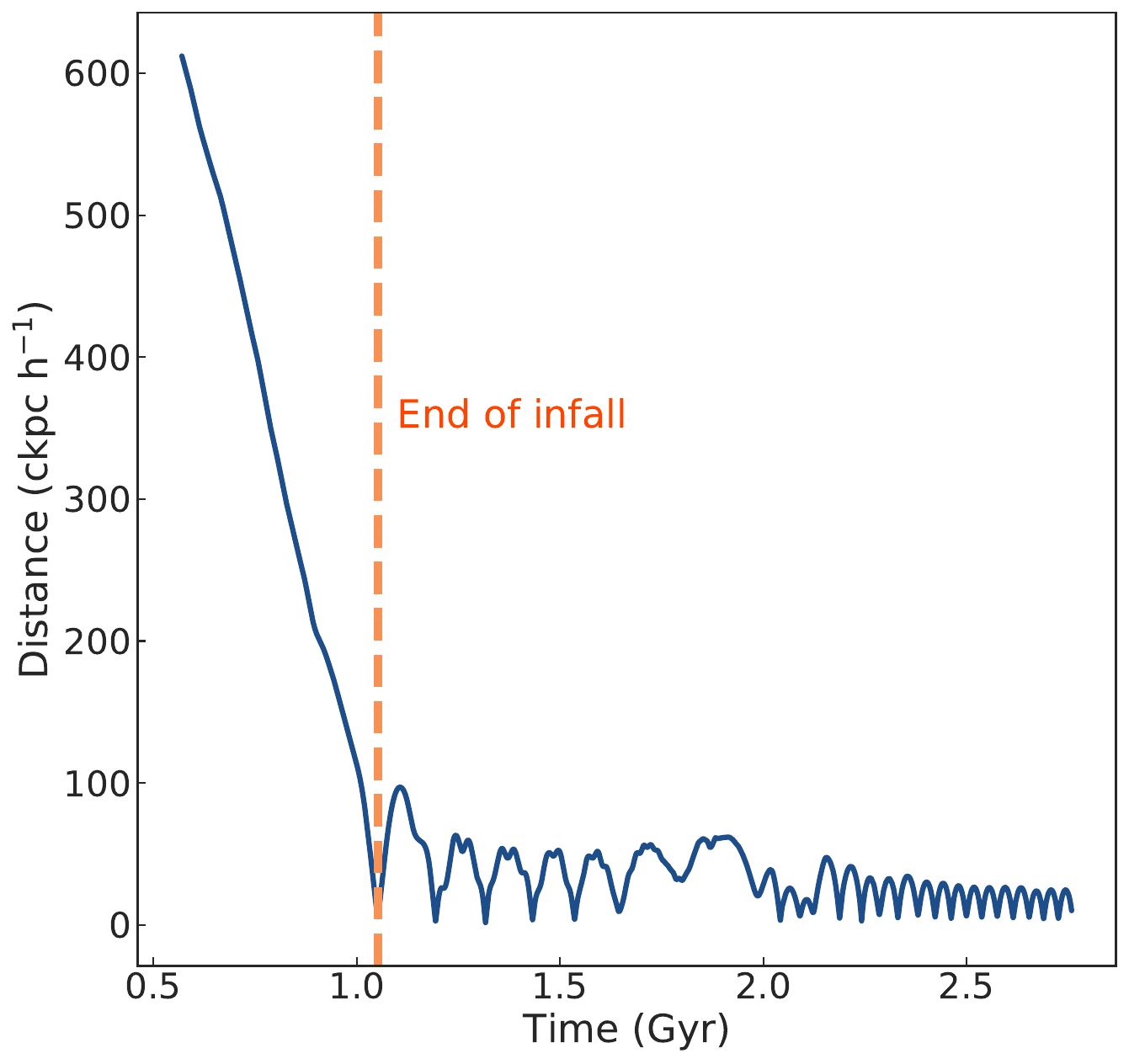}
    \caption{Galactocentric distance vs. time for the reference IMBH, with respect to its galaxy's central SMBH. The vertical line marks the end of the infall, a period characterized by a galaxy-galaxy merger between hosts. The periodic figures in the plot indicate orbits in the galactocentric frame.}
    \label{fig:infall_demo}
\end{figure}

\subsection{Galaxy selection} \label{sec:galaxy_select}

Structures in \astrid are divided into FOF groups (halos), which are further divided into subgroups (also called subhalos). We studied the groups at $z = 3$, which was the lowest redshift available at the time we performed this study.

With the halo-based BH seeding algorithm applied in \texttt{ASTRID}, the population of wandering BHs are mostly brought by the galaxy mergers through the hierarchical structure formation.
The mean occupation number of the BHs grows with their host galaxy mass \cite[see][]{Ni_2022}.
To study the population of wandering BHs in massive galaxies at $z \sim 3$, we selected three examples of massive halos at this redshift. For each halo, we selected the subgroup that contains the most BHs (which also has the most massive galaxies). We refer to these three subgroups as Galaxies 1, 2, and 3, with their illustration shown in Fig. \ref{fig:galaxies} and physical properties listed in Table \ref{tab:galaxy_properties}.

\subsection{Black hole selection} \label{sec:BH_select}

We considered the largest BH in each galaxy to be the central SMBH. The central SMBHs in Galaxies 1, 2, and 3 are marked in Fig. \ref{fig:galaxies}, and their masses are given in Table \ref{tab:galaxy_properties}. We refer to the non-central BHs as wandering BHs. Note that the central SMBH is characterized as being several orders of magnitude more massive than all the other BHs in the galaxy. In each of the three studied galaxies, our selected central SMBH is the closest BH of this category to the center-of-mass. These central SMBHs can be slightly shifted with respect to the center-of-mass of the stellar distribution by a few kiloparsecs; this is due to the very asymmetric stellar distributions characteristic of the galaxies studied.

For each wandering BH, we calculated the distance from the galaxy's central SMBH at every time point. Any wandering BH with a mean galactocentric distance under $100 \dist$ was selected to be studied in detail. As reported in Table \ref{tab:galaxy_properties}, we found eight IMBHs that meet this requirement in Galaxy 1, twelve in Galaxy 2, and eight in Galaxy 3, for a total of $28$ IMBHs. These IMBHs, marked in Fig. \ref{fig:galaxies}, have masses ranging from $\sim 4.7 \times 10^4 \Msun$ to $\sim 2.7 \times 10^6 \Msun$ at $z = 3$. We note that this covers most of the intermediate-mass range, in contrast to our previous work with the Illustris TNG50 simulation, in which we could only investigate small IMBHs due to our use of stellar clusters as proxies \citep{Weller_2022}.

\begin{figure}
	\includegraphics[width=\columnwidth]{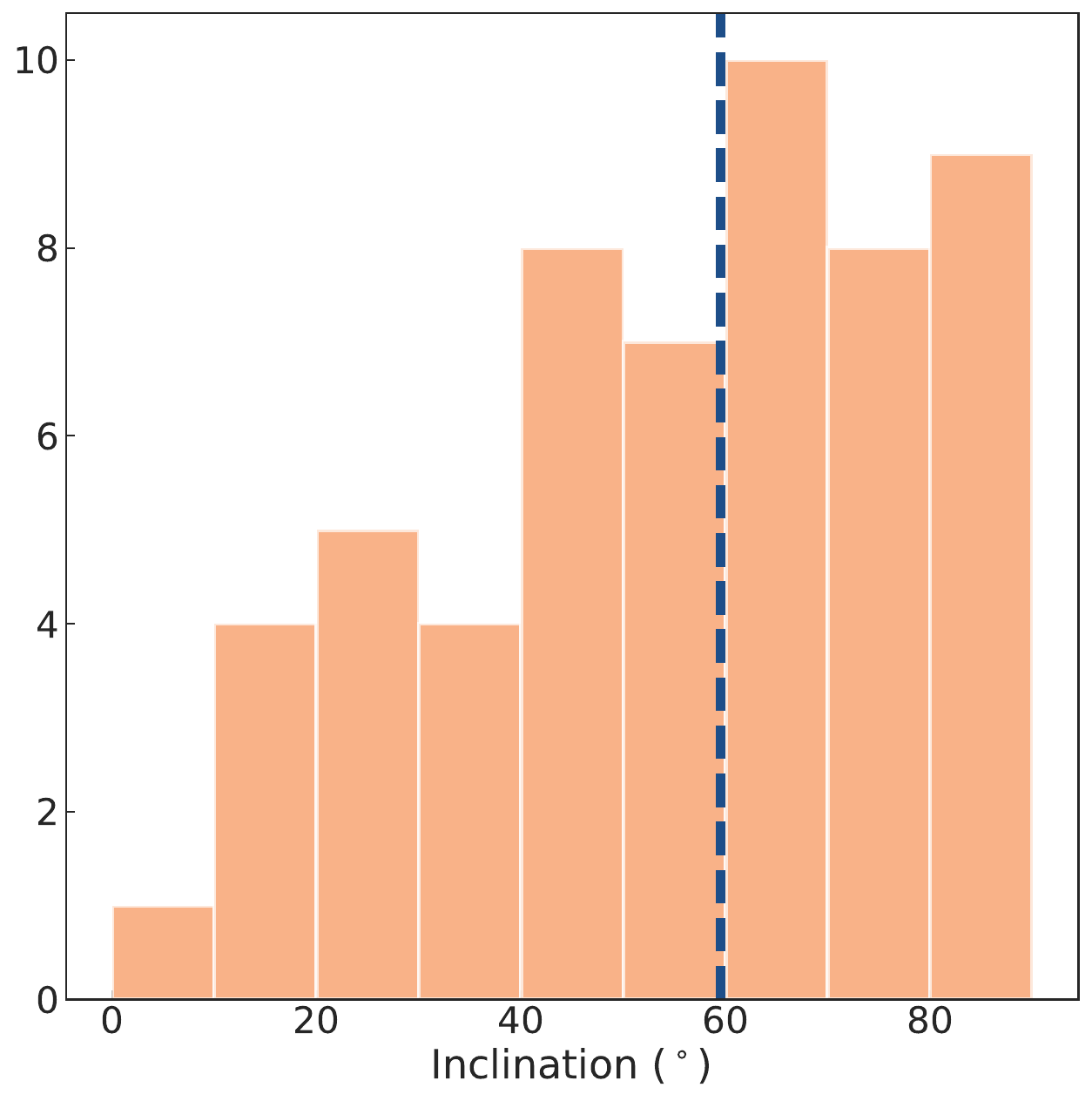}
    \caption{Distribution of orbital inclinations of our sample of wandering IMBHs in massive galaxies at $z \sim 3$. Large inclination values are evident, with a median value of $60^\circ \pm 22^\circ$.}
    \label{fig:inclination_histogram}
\end{figure}

\begin{figure}
	\includegraphics[width=\columnwidth]{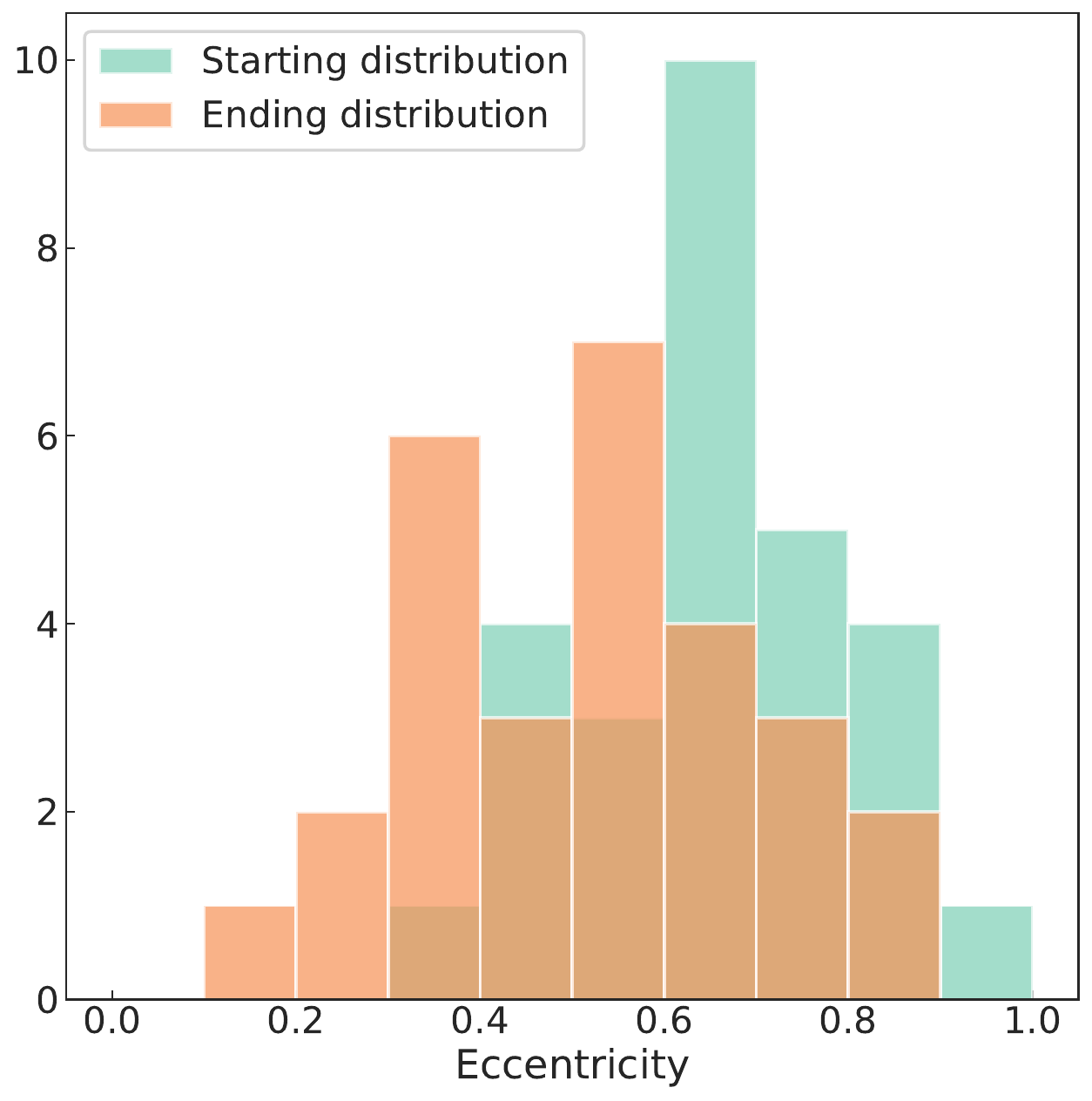}
    \caption{Histograms of the mean eccentricity over the first three and last three orbits of each of the 28 studied BHs. The ending distribution has a lower mean, indicating that the orbits become less eccentric with time.}
    \label{fig:ecc_hist}
\end{figure}

\section{Orbital Properties and Evolution} \label{sec:orbits}
In this Section, we discuss some properties, such as orbital inclination and eccentricity, and how they evolve with time.

\subsection{Infall and orbits} \label{sec:infall}
A plot of the galactocentric distance vs. time of wandering IMBHs shows an initial infall followed by a periodic behavior. An example is shown in Fig. \ref{fig:infall_demo} for one of the studied IMBHs from Galaxy 3; we label this object as the reference IMBH from now on. The reference IMBH was chosen because its mass, $6.7\times 10^4 \Msun$ at $z=3$, is close to the median mass of IMBHs investigated.
The infall period corresponds to the IMBH entering the galaxy, likely during a galaxy-galaxy merger. The shape of the periodic figure in the distance vs. time plot suggests eccentric orbits.
We use the term ``post-infall'' to refer to snapshots of an IMBH after the galactic merger's end. We used the beginning of the first period to mark the end of the infall (see Fig. \ref{fig:infall_demo}). We prefer this empirical approach rather than relying on the subhalo's merging time because the orbits are very well defined by the distance vs. time relations. However, the time of the merger event can be slightly different from what we assess by using this methodology.

Orbital identification and separation are prerequisites to studying the evolution of orbital properties. For each studied IMBH, we chose a distance $R = D$ around the middle of the distance range covered by the orbital period. Whenever we had $R < D$ at one snapshot but $R > D$ at the following snapshot, we marked the second snapshot as the beginning of the orbit.

\subsection{Orbital plane inclination} \label{sec:galactic_plane}

\begin{figure*}
	\includegraphics[width=2\columnwidth]{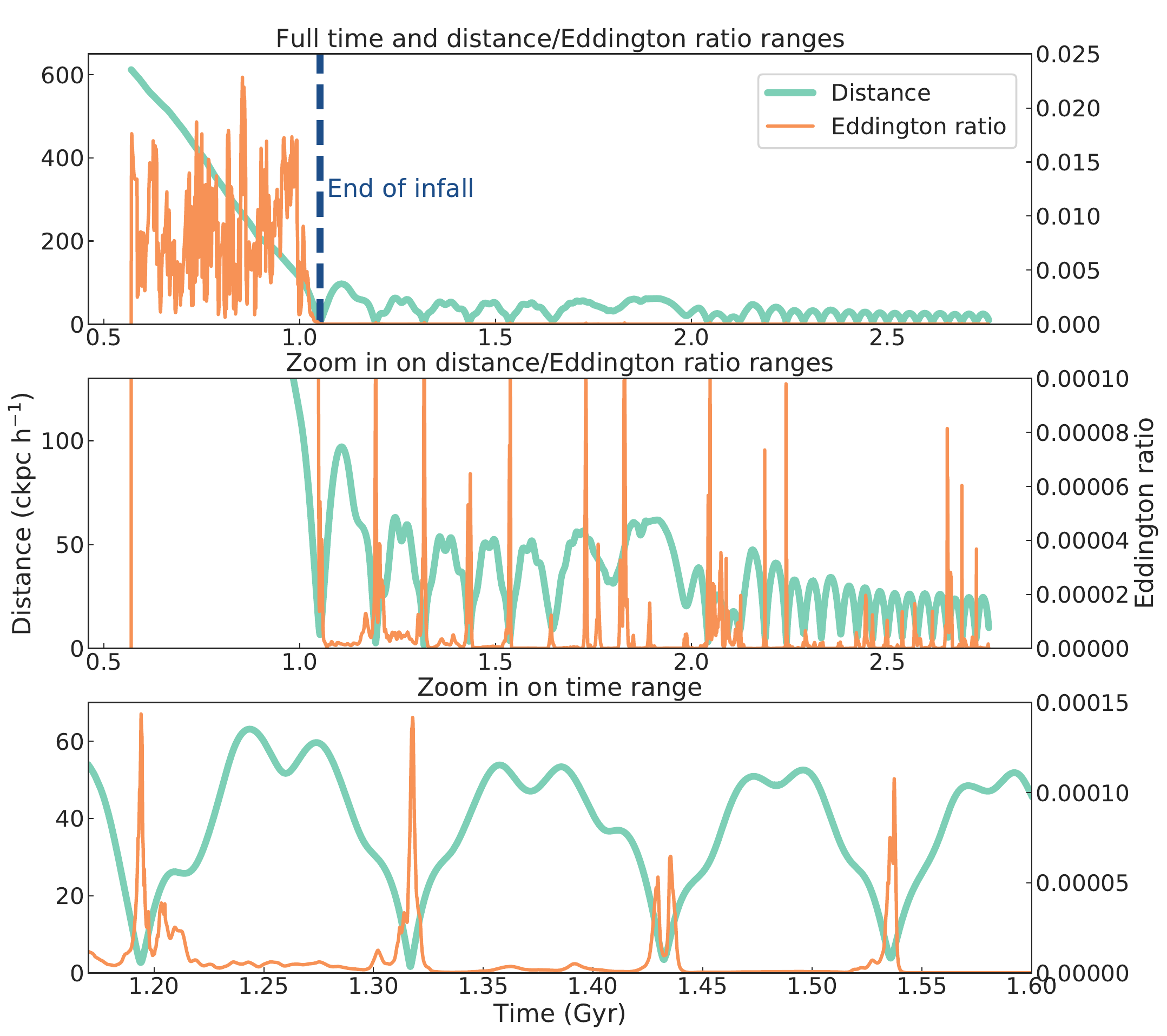}
    \caption{Galactocentric distance (thick, green line) and Eddington ratio (thin, orange line) vs. time for the reference IMBH. In the top panel, the Eddington ratio spikes high during the infall period. In the middle panel, the vertical axes are cut for better visualization, and the Eddington ratio spikes when the distance is close to a minimum. In the bottom panel, we zoom in to a small time range section to observe the spikes more clearly.}
    \label{fig:mdot_spikes}
\end{figure*}

Next, we studied the inclination in our sample of IMBHs relative to the galactic plane. To find the orientation of the orbital plane, we used the IMBH position $\vec R$ and velocity $\vec V$ relative to the central SMBH. To find the orientation of the main galactic plane, we used the SubhaloSpin property, which provides the angular momentum vector $\vec L$ of the galaxy. Note that the SubhaloSpin property is provided at low time resolution, so we calculate orbital inclinations only at specific times when this information is available — namely $z = 3$ and $z = 4$.

Then, to find the orbital inclination $i$, we first calculated $\vec C = \vec R \times \vec V$, which is perpendicular to the orbital plane. Then $i$ is the angle between $\vec C$ and $\vec L$, considered to be less or equal to $90^\circ$:
\begin{equation}
    i = \arccos{\frac{\vec C \cdot \vec L}{\lVert \vec C \rVert \, \lVert \vec L \rVert}} \, .
\end{equation}

We display in Fig. \ref{fig:inclination_histogram} the distribution of orbital inclinations for our sample of IMBHs in massive $z \sim 3$ galaxies. Our analysis shows that wandering IMBHs do not typically belong to the main plane of their host galaxies. Instead, inclinations are quite large, with a median value of $\sim 60^\circ$ and a standard deviation of $\sim 22^\circ$ over the distribution. 
High-inclination IMBHs interact with the host galaxy's plane for the shortest time, hence they experience very little dynamical friction. Consequently, IMBH-IMBH mergers involving high-inclination objects are highly unlikely.
Large orbital inclinations can cause lower duty cycles of accretion, as the IMBH spends only a fraction of time within the galactic disk, where the density of the interstellar medium (ISM) is high (see \S \ref{sec:edd_ratio}).

\subsection{Orbital eccentricity} \label{sec:period-ecc}
We calculated the orbital eccentricity as $e = (R_a - R_p)/(R_a + R_p)$, where $R_a$ and $R_p$ are the apocenter and pericenter distances, respectively. 
As explained in \S \ref{sec:BH_select}, for this study we select only IMBHs with a mean galactocentric distance under $100 \dist$. We caution that this criterion may eliminate the IMBHs with the most eccentric orbits.

In our sample, the median eccentricity over all orbits of our 28 studied IMBHs is $\sim 0.6$, with a standard deviation of $\sim 0.2$. 
We also displayed in Fig. \ref{fig:ecc_hist} the mean eccentricity at the first three and last three orbits of each IMBH — the mean eccentricity decreased for $\sim 71\%$ of our studied population. The ending distribution has a mean of $e \sim 0.5$, while the starting distribution's mean is $e \sim 0.7$. This supports the conclusion that orbits of wandering IMBHs tend to become less eccentric with time \citep{Weller_2022}.

\section{Accretion and radiation} \label{sec:accretion}
In this final Section, we discuss the accretion and radiative properties of wandering IMBHs in massive galaxies at $z \sim 3$. We estimate typical accretion rates, duty cycles, X-ray luminosity, and detectability.

\subsection{Eddington ratio} \label{sec:edd_ratio}
We calculated the Eddington ratio $f_{\rm Edd} \equiv \dot M_\bullet/\dot M_{\rm Edd}$, i.e., the ratio between the accretion rate and its Eddington value:
\begin{equation}
    \dot M_{\rm Edd} =  2.2 \times 10^{-8} \left( \frac{\Mblack}{\Msun} \right) \rm \, [\Msun yr^{-1}] \, ,
\end{equation}
assuming a standard value of $10\%$ for the matter-to-energy conversion efficiency \citep{Shakura1973}.
The values of the black hole mass and accretion rate are retrieved directly from the simulation.

We are interested in investigating how the accretion rate depends on the location of the IMBH. As orbits are very eccentric, we consider the relation between the galactocentric distance of the IMBH and its Eddington ratio.
An analysis of the time evolution of the Eddington ratio and galactocentric distance of individual IMBHs  shows spikes in the accretion rate when the distance is low (see Fig. \ref{fig:mdot_spikes}). This fact suggests that wandering IMBHs undergo spikes in accretion activity only when sufficiently close to the galactic center, where the ISM density is high. During these events, the median maximum Eddington ratios reached are of the order of $10^{-3}$ — about one order of magnitude higher than the Eddington ratios reached by our reference IMBH during its activity spikes.

In the top panel of Fig. \ref{fig:mdot_spikes}, the Eddington ratio spikes to $\sim 0.02$ during the infall period.
The significant accretion rate during the infall phase is a typical pattern for all the wandering IMBHs investigated.
This period of high Eddington rates signals that the IMBH originally resided at the center of its host galaxy before the galactic encounter. Hence, during the galactic merger — likely a minor one — the gas of its host was tidally stripped, and the IMBH started to wander around with low Eddington rates. \cite{Steinborn_2015}, in this regard, show that the gas of the secondary galaxy is rapidly accreted onto the central SMBH of the main galaxy.

The trend of Eddington ratios increasing with decreasing distance from the galactic center is typical for all IMBHs in our sample.
This fact has profound repercussions for the duty cycle of sources powered by accretion onto wandering IMBHs. In general, the duty cycle is the fraction of the time that an IMBH spends accreting.
We estimate the duty cycle in the post-infall phase from Fig. \ref{fig:mdot_spikes}.
We first calculate the median value of the Eddington ratio over the post-infall time range.
The median value represents the ground floor accretion level, which, in the case of the reference IMBH, turns out to be $f_{\rm Edd} \sim 10^{-6}$.
Then, we calculate the fraction of time during which the Eddington ratio is at least ten times larger than this floor value.
We obtain a duty cycle of ${\cal D} \sim 13\%$. Repeating this calculation for all of our studied IMBHs, we find a median of ${\cal D} \sim 12 \%$.
This value of the duty cycle for accretion indicates that, although many IMBHs may wander nearby massive galaxies, they would be accreting at non-negligible levels only for a small fraction of time. As orbits become less eccentric and IMBHs tend to wander closer to the galactic center \citep{Weller_2022}, we expect the duty cycle to increase with cosmic time.

\subsection{X-ray luminosity, detectability, and spectral energy distribution} 
\label{sec:detectability}
Lastly, we make predictions for the $2-10$ keV X-ray luminosity and detectability of wandering IMBHs in massive $z \sim 3$ galaxies, and calculate the typical spectral energy distribution (SED).

\begin{figure}
	\includegraphics[width=\columnwidth]{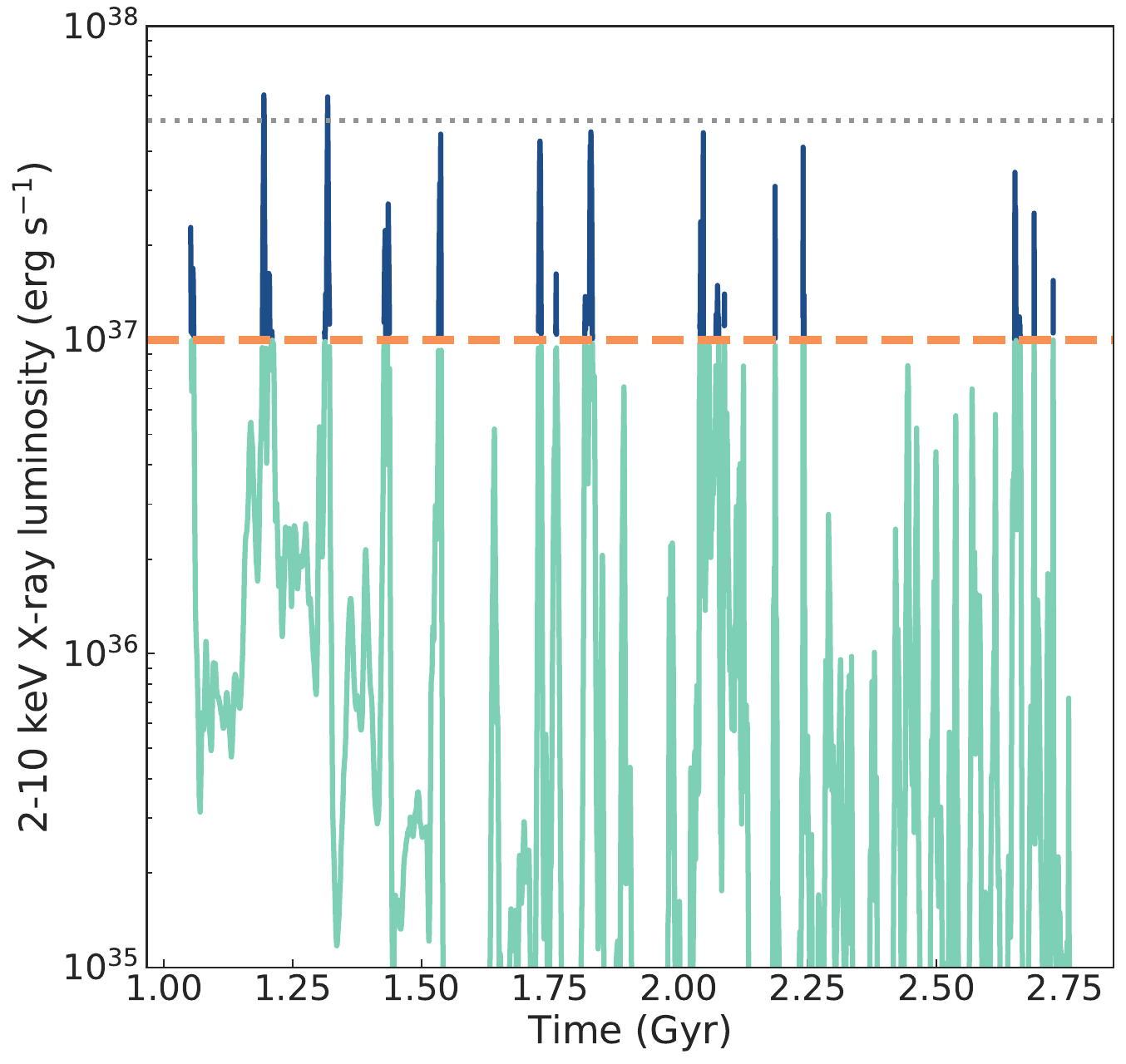}
    \caption{X-ray luminosity in the $2-10$ keV band produced by our reference IMBH. The thick, dashed, orange line demarcates the luminosity threshold of $10^{37} \, \mathrm{erg \, s^{-1}}$, which corresponds to fluxes of $10^{-15} \rm \, erg \, s^{-1} \, cm^{-2}$ within $10$ Mpc, observable by Chandra with integration times of $\sim 200$ ks. The thin, dotted, gray line corresponds to the level of X-ray background emission (from XRBs and the ISM) computed from the SFR within $20 \dist$ from the reference IMBH at $z = 3$. Note that there is no active star formation at $z = 3$ within $10 \dist$.}
    \label{fig:fx_spikes}
\end{figure}

For the reference IMBH in Fig. \ref{fig:mdot_spikes}, we calculated the $2-10$ keV X-ray luminosity from the Eddington ratio, assuming an X-ray bolometric correction of $5\%$ \citep{Vasudevan_2007, Vasudevan_2009}.
The result is shown in Fig. \ref{fig:fx_spikes}.
The X-ray luminosity of $10^{37} \, \mathrm{erg \, s^{-1}}$ is marked as a threshold to indicate the highest luminosity spikes. An X-ray luminosity of $10^{37} \, \mathrm{erg \, s^{-1}}$ corresponds to an X-ray flux of $10^{-15} \, \mathrm{erg \, s^{-1} \, cm^{-2}}$ if the source is at $10$ Mpc — observable by the Chandra X-ray observatory with integration times of $\sim 200$ ks, according to the Chandra proposal planning tools.
The X-ray detection fraction is estimated as ${\cal D}_x \sim 4\%$ for this particular reference IMBH. Repeating this calculation for all of our studied IMBHs, we find a median of ${\cal D}_x \sim 12\%$.

As explained in \S \ref{sec:ASTRID}, we use the standard value of $10\%$ for the matter-to-energy radiative efficiency. For meager Eddington rates, the accretion flow becomes radiatively inefficient (see, e.g., \citealt{Rees_1982, Blandford_1999, Quataert_2000}); some models argue for a different conversion between accretion rate and luminosity (see, e.g., \citealt{Mineshige_2000, Churazov_2005, Dangelo_2015}). For example, \cite{Churazov_2005} proposes a model in which the bolometric luminosity emitted in the radiatively inefficient regime, for $f_{\rm Edd} < 0.1$, is $10 f_{\rm Edd}$ times lower than in the radiatively efficient case — making our IMBHs undetectable, as they are characterized by typical Eddington ratios $f_{\rm Edd} < 10^{-5}$.

Depending on the level of star formation rate (SFR) in their hosts, these sources may be hard to disentangle from other X-ray sources, e.g., X-ray binaries (XRBs) and emission from the hot ISM.
Recent studies by \cite{Koudmani_2021}, \cite{Haidar_2022} and \cite{Sharma_2022} perform a careful analysis of how the X-ray emission of the wandering massive BHs population compares with other X-ray emitters, although for galaxies that are $\sim 100$ times less massive than those we have considered in this study.
In particular, they use relations connecting the host's stellar mass and SFR with the emission from XRBs \citep{Lehmer_2019} and the hot ISM \citep{Mineo_2012_ISM}.
We use the relation in \cite{Lehmer_2010} between the $2-10$ keV luminosity from XRBs and the SFR in our range of interest
\begin{equation}
   L^{\rm XRB}_{X} \, [\mathrm{erg \, s^{-1}}] = 10^{39.57} \times \mathrm{SFR}^{0.94} \, .
\end{equation}
Similarly, we use the relation
\begin{equation}
   L^{\rm ISM}_{X} \, [\mathrm{erg \, s^{-1}}] = 8.3 \times 10^{38} \times \mathrm{SFR} \, ,
\end{equation}
for the ISM component from \cite{Mineo_2012_ISM}.
If the host galaxy is spatially unresolved, its wandering IMBH population is likely undetectable. In fact, the total SFRs of typical massive galaxies in our sample are very large, i.e., $\mathrm{SFR} > 1000 \, \mathrm{\Msun \, yr^{-1}}$; hence, their X-ray emission would largely overcome that from the IMBHs.
In the case of spatially resolved galaxies, the situation is different — some wandering IMBHs could be located in regions of low background X-ray emission, especially at large galactocentric distances. In the case of the reference IMBH, whose X-ray light curve is shown in Fig. \ref{fig:fx_spikes}, we find no active star formation at $z=3$ within $10 \dist$, while the SFR within $20 \dist$ is $\sim 0.01 \, \mathrm{\Msun \, yr^{-1}}$. With this level of SFR, we find $L^{\rm ISM}_{X} \sim  10^{37} \, \mathrm{erg \, s^{-1}}$, and $L^{\rm XRB}_{X} \sim 5 \times 10^{37} \, \mathrm{erg \, s^{-1}}$. These values are comparable to the highest spikes in the X-ray luminosity we predict, as shown in Fig. \ref{fig:fx_spikes}. We conclude that, in the case of spatially resolved hosts, the detectability of this wandering IMBH population depends on the star-forming activity in their immediate neighborhood. It is conceivable that the highest spikes in X-ray luminosity coincide with passages in high-density regions where the SFR is also large.

We note that some IMBHs in our sample are characterized by very high spikes of X-ray luminosity, reaching levels of $10^{41} \, \mathrm{erg \, s^{-1}}$. These luminosities are typical of hyper-luminous X-ray sources (HLXs). Recently, \cite{Barrows_2019} selected a sample of $\sim 20$ HLX candidates in the Chandra Source Catalog. These sources are off-nuclear in galaxies up to a redshift $z \sim 0.1$ and are consistent with a mass in the IMBH range. \cite{Mezcua_2018_HLX} select a sample of HLXs up to $z \sim 2.4$. In our work, we find that $2$ out of the $28$ IMBHs studied ($\sim 7 \%$) reach the luminosity of HLXs. In particular, one IMBH spikes in the HLX-luminosity level only for a time $\sim 0.5$ Myr, while the other one keeps this luminosity level for $>150$ Myr as it inspirals towards the center of the galaxy and eventually merges with the central SMBH. This suggests that HLXs are a small subset of the general wandering IMBH population, with typical X-ray luminosities $\sim 3-4$ orders of magnitude lower. These IMBHs reaching HLX luminosities are those that are more massive and have more significant accretion rates because they sink deeper into the galactic potential.

\begin{figure*}
	\includegraphics[width=2\columnwidth]{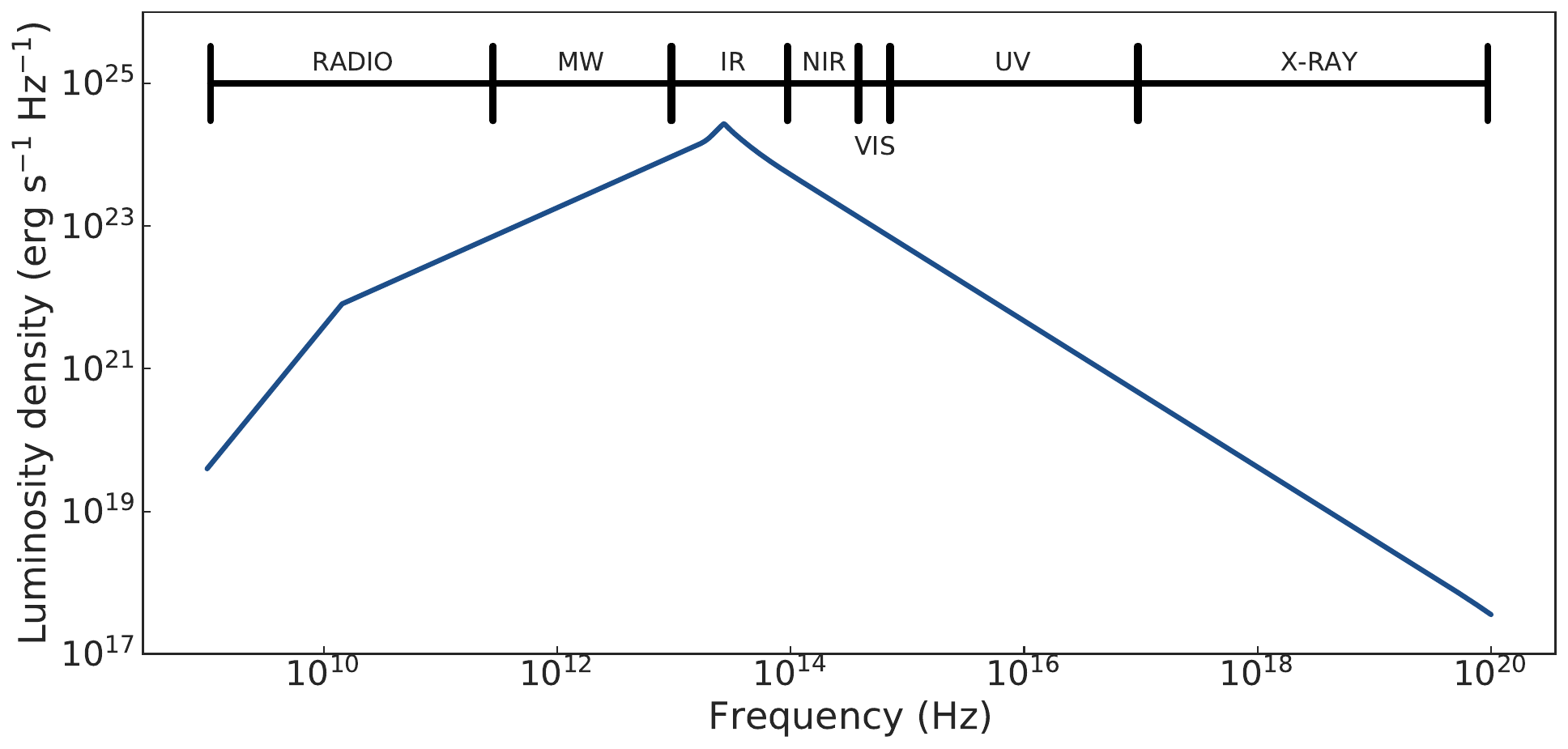}
    \caption{Typical rest-frame SED of an IMBH in our study, calculated using the mean mass and Eddington ratio of our sample.}
    \label{fig:h0_SED}
\end{figure*}

We also calculated a typical SED for a wandering IMBH passing within the inner parts of a galactic disk.
We used a code developed for black holes accreting in advection-dominated accretion flows (ADAF) mode, i.e., for strongly sub-Eddington rates \citep{Pesce_2021, Seepaul_2022}. In this simple analytical model, which is based on the original study by \cite{Mahadevan_1997}, the emission results from a combination of synchrotron, bremsstrahlung, and inverse
Compton radiation. Note that this treatment does not include the jet component of the emission. In this analytical model, the SED depends only on two parameters: black hole mass and Eddington ratio. The velocity of the IMBH with respect to the local gas frame enters the calculation of the SED only indirectly, i.e., through the determination of the accretion rate onto the IMBH.

To produce this ``typical'' SED, we calculated the mean mass and Eddington ratio over all time points of the 28 studied BHs corresponding to galactocentric distances under $2 \dist$. The resulting SED is shown in Fig. \ref{fig:h0_SED}, peaking in the infrared at a rest-frame frequency of $\sim 2.7 \times 10^{13} \rm \, Hz$, or $\sim 11 \, \mathrm{\mu m}$, with a maximum luminosity density of $\sim 2.7 \times 10^{24} \rm \, erg \, s^{-1} \, Hz^{-1}$.

\section{Discussion and Conclusions} \label{sec:conclusions}
The astronomical community is growing increasingly aware that a population of IMBHs, with masses between $10^3 \Msun$ and $10^6 \Msun$, might be wandering, thus far largely undetected, within the volume of galaxies. These IMBHs might form through two avenues: (i) in-situ formation via gravitational runaway in star clusters (see, e.g., \citealt{HB_2008, Komossa_Merritt_2008, Fragione_2018a}), or (ii) ex-situ formation at the center of a dwarf galaxy (see, e.g., \citealt{Governato_1994, Schneider_2002, OL_2009, Greene_2020_IMBH, Greene_2021}).

Our work was motivated by our previous analysis of wandering IMBHs in the Illustris TNG50 simulation \citep{Weller_2022}, for which we had to resort to stellar clusters as proxies to investigate IMBHs, due to the lack of BHs in the appropriate mass range.
In the cosmological simulation \astrid \citep{Ni_2022, Bird_2022}, BHs in the mass range of IMBHs are self-consistently modeled, easing our way to investigate their orbital and radiative properties in massive galaxies at $z \sim 3$. We caution the reader that these galaxies are more gas-rich than local ones. In fact, at redshifts lower than $z \sim 1$ the amount of cold gas available for IMBH accretion steadily declines until $z = 0$ \citep{Power_2010}, although \cite{Pacucci_2020} find that, in the local Universe, black holes with a mass $\Mblack < 10^8 \Msun$ still grow mostly by accretion, as opposed to by mergers.
For these reasons, the predictions we make in this study do not immediately translate for $z=0$ IMBHs.

Our main results are as follows:
\begin{itemize}
    \item Wandering IMBHs have \textbf{large orbital inclinations} with respect to the principal plane of their host. The median orbital inclination in our sample is $60^\circ \pm 22^\circ$. Some IMBHs have orbits perpendicular to the galactic plane.
    \item Wandering IMBHs have \textbf{large orbital eccentricities}, with a median of $0.6 \pm 0.2$. Eccentricities slightly decrease with time.
    \item Typical \textbf{accretion duty cycles are low} for wandering IMBHs, with a median of $\sim 12 \%$ — a cycle of accretion activity commences when, during their orbit, IMBHs approach the pericenter.
    \item IMBHs in massive galaxies have \textbf{spikes of accretion activity}, during which they can be detected in the X-rays. They reach $2-10$ keV X-ray luminosities $>10^{37} \, \mathrm{erg \, s^{-1}}$ during $\sim 10\%$ of the timeframe investigated. This corresponds to X-ray fluxes $>10^{-15} \, \mathrm{erg \, s^{-1} \, cm^{-2}}$, which are detectable within $10$ Mpc by Chandra with exposures of $\sim 200$ ks.
    \item Two out of the 28 IMBHs studied
    ($\sim 7\%$) reach \textbf{X-ray luminosities $> 10^{41} \, \mathrm{erg \, s^{-1}}$, in the HLXs regime.} HLXs are a small subset of the general wandering IMBHs population, characterized by luminosities $10^3 - 10^4$ times fainter. 
    \item Typical wandering IMBHs have \textbf{SEDs with a peak in the infrared band}, at $\sim 2.7 \times 10^{13} \rm \, Hz$, or $\sim 11 \, \rm \mu m$ rest frame.
\end{itemize}

Our results suggest that wandering IMBHs have eccentric and very tilted orbits, and thus they accrete at levels of $10^{-3} - 10^{-5}$ Eddington only near the pericenter of their orbits. Nonetheless, during these brief outbursts of accretion activity, their X-ray signature could be detectable with modest integration times by current and future X-ray facilities. Extensive surveys are needed to assess the demographics of the population IMBHs in galaxies.

\section*{Acknowledgements} \label{sec:acknowledgements}
E.W. acknowledges undergraduate research support provided by the Harvard College Research Program (HCRP).
F.P. acknowledges support from a Clay Fellowship administered by the Smithsonian Astrophysical Observatory. This work was also supported by the Black Hole Initiative at Harvard University, which is funded by grants from the John Templeton Foundation and the Gordon and Betty Moore Foundation.
Y.N. acknowledges support from the ITC fellowship from Center for Astrophysics | Harvard \& Smithsonian.
T.D.M. acknowledges funding from the NSF AI Institute: Physics of the Future, NSF PHY-2020295, NASA ATP NNX17AK56G, NASA ATP 80NSSC18K101, and NASA ATP 80NSSC20K0519.
\astrid is carried out on the Frontera facility at the Texas Advanced Computing Center. 
\section*{Data Availability} 
Part of the \astrid data are available at \url{https://astrid-portal.psc.edu/}.
The codes used to analyze the data will be shared on reasonable request to the corresponding author.



\bibliographystyle{mnras}
\bibliography{ms} 





\bsp	
\label{lastpage}
\end{document}